\documentclass[preprint,12pt]{elsarticle}

\usepackage{amssymb}
\usepackage{color}

\journal{Physica A}

\begin{document}

\begin{frontmatter}



\title{Coupled effects of local movement and global interaction on contagion}


\author{Li-Xin Zhong$^{a,b}$}\ead{zlxxwj@163.com}
\author {Wen-Juan Xu$^c$}
\author {Rong-Da Chen$^a$}
\author {Tian Qiu$^d$}
\author {Yong-Dong Shi$^e$}
\author{Chen-Yang Zhong$^f$}

\address[label1]{School of Finance and Coordinated Innovation Center of Wealth Management and Quantitative Investment, Zhejiang University of Finance and Economics, Hangzhou, 310018, China}
\address[label2]{School of Economics and Management, Tsinghua University, Beijing, 100084, China}
\address[label3]{School of Law, Zhejiang University of Finance and Economics, Hangzhou, 310018, China}
\address[label4]{School of Information Engineering, Nanchang Hangkong University, Nanchang, 330063, China}
\address[label5]{Research Center of Applied Finance, Dongbei University of Finance and Economics, Dalian, 116025, China}
\address[label6]{Yuanpei College, Peking University, Beijing, 100871, China}

\begin{abstract}
By incorporating segregated spatial domain and individual-based linkage into the SIS (susceptible-infected-susceptible) model, we propose a generalized epidemic model which can change from the territorial epidemic model to the networked epidemic model. The role of the individual-based linkage between different spatial domains is investigated. As we adjust the timescale parameter $\tau$ from 0 to unity, which represents the degree of activation of the individual-based linkage, three regions are found. Within the region of $0<\tau<0.02$, the epidemic is determined by local movement and is sensitive to the timescale $\tau$. Within the region of $0.02<\tau<0.5$, the epidemic is insensitive to the timescale $\tau$. Within the region of $0.5<\tau<1$, the outbreak of the epidemic is determined by the structure of the individual-based linkage. As we keep an eye on the first region, the role of activating the individual-based linkage in the present model is similar to the role of the shortcuts in the two-dimensional small world network. Only activating a small number of the individual-based linkage can prompt the outbreak of the epidemic globally. The role of narrowing segregated spatial domain and reducing mobility in epidemic control is checked. These two measures are found to be conducive to curbing the spread of infectious disease only when the global interaction is suppressed. A log-log relation between the change in the number of infected individuals and the timescale $\tau$ is found. By calculating the epidemic threshold and the mean first encounter time, we heuristically analyze the microscopic characteristics of the propagation of the epidemic in the present model.
\end{abstract}


\end{frontmatter}


\section{Introduction}
\label{sec:introduction}
Over the past two decades, the world has witnessed the threat of infectious disease with long-range transmission, such as the severe acute respiratory syndrome (SARS), the pandemic Influenza A (H1N1) and the 2014 Ebola epidemic. An outbreak of the epidemic in human society is closely related to the spatial scale and the interaction structure. G. Chowell et al. have investigated the growth patterns of the 2014 Ebola epidemic at different spatial scales\cite{chowell}. They have found that the global and the local growth rates exhibit different scaling laws. Depending upon the Cuban contact-tracing detection system, S. Clemencon et al. have studied the HIV/AIDS epidemics in Cuba\cite{clemencon}. They have found that the contact network exhibits a pattern with high intra-connectivity and low inter-connectivity. Understanding the reaction-diffusion processes in complex systems and finding an effective way to prevent and control the outbreak of infectious disease have become an urgent but difficult challenge\cite{tang1}.

In modelling the propagation of infectious disease, the susceptible-infected-susceptible (SIS) model and the susceptible-infected-recovered (SIR) model represent two distinctive disease transmission processes\cite{zhong1,boguna,mieghem}. In the SIS model, each individual is initially in one of the two states: S (susceptible) and I (infected). A susceptible individual may become infected as he encounters an infected individual, and an infected individual may become susceptible again after some time. In the SIR model, each individual is initially in one of the three states: S, I and R (recovered). The only difference between the SIS model and the SIR model is that, in the SIR model, an infected individual firstly becomes healthy (recovered), when he is immune to the disease. After some time, the recovered individual may become susceptible to the disease again.

The spread of infectious disease is closely related to mobility patterns and interaction patterns of the population\cite{liu1,graser,gong,schweitzer2}. An individual may move about aimlessly like a floating boat in the ocean or travel along a certain path like a move on the chessboard. Two individuals may interact with each other day after day because they are in the same firm or only have a chance encounter because neither of them likes establishing a fixed relationship with any other people. The mobility patterns and the interaction patterns have been extensively studied in the coevolutionary dynamics. M. Perc et al. have studied the coevolution of individual strategies and interaction structures, the coevolutionary rules are quite important in the occurrence of a variety of dynamical effects\cite{perc10}. A. Szolnoki et al. have studied the effects of mobility patterns on the emergence of cyclic dominance, the mutual relations of pattern formation, the impact of mobility and the emergence of cyclic dominance are found\cite{szolnoki10}. The original research related to the promising field can be found in refs\cite{perc11,perc12,szolnoki11}. In order to understand the of underlying structure in the evolution of collective behaviors, different network models are usually borrowed to simulate the mobility patterns and the interaction patterns\cite{hong,belik,wang1,apolloni,crepey,iribarren,tang}. The existing network models can be classified into two families: the network with a single layer, including regular network, random network and scale-free network\cite{lopez,moore,pastor,serrano,altarelli}, and the network with multiple-layers\cite{zheng10,shao2,gao,wang2,saumell}, which consists of two or more sub-networks. S. V. Buldyrev et al. have studied the properties of two interdependent networks\cite{buldyrev10}, a broader degree distribution is found to increase the vulnerability of the interdependent networks. J. X. Gao et al. have studied the percolation properties of interacting networks\cite{gao10}, an analytical framework has been developed.

In relation to the role of network structures in the outbreak of the epidemic, Z.H. Liu et al. have investigated the role of community structure in epidemic propagation\cite{liu2,wu,liu4}. They have found that the existence of community structure leads to a smaller threshold of epidemic outbreak. Z.M. Yang et al. have studied the effect of connection patterns on the outbreak of epidemic. They have found that a heterogeneous network structure accelerates the spread of infectious disease\cite{yang2,lu,hwang,hu2}. K. P. Chan et al. have investigated the effect of aging and links removal on epidemic spread in scale-free networks. The local and global spreads are found to be related to the number of links removed\cite{chan}.  M. Dickison et al. have studied the epidemic spreading on interconnected networks. A global endemic state is found to be related to the strongly-coupled or weakly-coupled network systems\cite{dickison}. X. F. Fu et al. have studied the effect of mobility patterns on the outbreak of epidemic. They have found that the change of encounter probability, which may result from the change of population density or the change of moving probability, plays an important role in the change of infected individuals\cite{fu2,yan}. V. Colizza et al. have investigated the invasion dynamics in metapopulation systems. They have found that the heterogeneous connectivity and the mixed mobility pattern play an important role in the outbreak of epidemic \cite{colizza}. C. Granell et al. have studied the interrelation between the spread of an epidemic and the risk awareness\cite{granell10}. They have found that there exists a critical point for the onset of the epidemics. M. Boguna et al. have studied the conditions for the absence of an epidemic threshold in heterogeneous networks\cite{boguna10}. A delicate balance between the number of nodes with a high degree and the topological distance is found to be quite important for the occurrence of the threshold.

Until today, in relation to the role of mobility patterns and connection patterns in the outbreak of epidemic, most of the studies consider the two factors respectively. However, in the real world, the evolutionary dynamics in reaction-diffusion systems is usually determined by the coupling of mobility and connectivity, especially the group dynamics in social and economic systems \cite{johnson1,johnson2,schweitzer1,perc1,fent}. For example, a susceptible individual may be infected by an unknown individual in a random walk after dinner or be infected by a business partner. The coupled effects of random walk and pre-arranged interaction on contagion are short of in-depth study. In relation to the measures for disease control and prevention, the government usually takes isolation measures to prevent the epidemic. For example, during the outbreak of the 2014 Ebola epidemic, most countries tighten up on the rules for admitting people into the country. Alvarez Zuzek et al. have measured the effect of isolation in the multiple network\cite{zuzek10}. The epidemic threshold is found to increase with the rise of the isolation time. S. Bonaccorsi et al. have studied the epidemic on the networks that are partitioned into local communities\cite{bonaccorsi10}. Their results indicate that knowing the spectral radius of the graph is quite important in judging whether an overall healthy state is stable or not. V.M. Kenkre et al. have studied the effect of confinement on the spread of epidemics in a system of random walkers who move in n dimensions\cite{kenkre10}. They have found that the spread of the epidemic has a non-monotonic dependence on the extent of the home range of the individuals. A. Apolloni et al. have studied the effect of population partitions, mixing patterns and mobility structures on the spread of the epidemic\cite{apolloni10}. Some efficient strategies controlling the pandemic potential have been found. The related work can also be found in refs\cite{bu10,poletto10,maeno10,potter10,lund10}. A. X. Cui et al. have studied the effects of strong ties on the spread of the epidemic\cite{cui10}. They have found that the epidemic prevalence is promoted by the strong ties. Although the mobility patterns and connection patterns have been considered as important factors for the spread of the epidemic, most of the studies focus on the effects of immunization and isolation, which especially rely on the knowledge of one's activity routine and one's friendship network. Generally speaking, compared with the route of random walk, a predefined interaction structure is much easier to be understood. In view of the coupling of random walk and predefined interaction structure, the effect of local activity restriction on global epidemic suppression and the role of the shortcuts existing between the individuals geographically far away are worth extra attention.

Inspired by the metapopulation epidemic model, the territorial epidemic model and the networked epidemic model in refs\cite{giuggioli,colizza10,belik10,lopez10,newman,kuperman10,satorras10,mieghem10}, in the present model, we incorporate a multi-layer network, which consists of a regular network with segregated spatial domains and a random network representing individual-based linkage, into the SIS model. The coupled dynamics of random walk and pre-arranged linkage is investigated. The role of narrowing segregated spatial domain and reducing mobility in epidemic control is checked. The motivation of the present work is to find out how the individual-based linkage between segregated spatial domains affects the outbreak of the epidemic globally, just as the role of the shortcuts in the outbreak of the epidemic on the small world network. The following are our main findings.

(1)Activating the prearranged linkage between different spatial domains can prompt the outbreak of the epidemic globally. As the value of timescale $\tau$ ranges continuously from 0 to unity, three regions are observed. Within the region of $0<\tau<0.02$, the epidemic is determined by local movement and is sensitive to the timescale $\tau$. Within the region of $0.02<\tau<0.5$, the epidemic is insensitive to the timescale $\tau$. Within the region of $0.5<\tau<1$, the outbreak of the epidemic is determined by the structure of the individual-based linkage. The role of activating the prearranged linkage in the present model is similar to the role of the shortcuts in the two-dimensional small world epidemic model. Only activating a small number of the prearranged linkage can effectively promote the global outbreak of the epidemic. 

(2)The size of the segregated spatial domain plays an important role in the outbreak of the epidemic. As both the overall spatial size and the population size are given, narrowing the segregated spatial domain not only slows down the spread of infectious disease but also leads to a decrease in the number of infected individuals in the final steady state.

(3)Local mobility has a great impact on the global spread of infectious disease. On condition that there is little individual-based connection, reducing local mobility can effectively suppress the global spread of epidemic.

(4)A heuristic analysis indicate that the change of the macroscopic characteristics of the propagation of the epidemic in the present model is closely related to the change of the scaling laws for the epidemic threshold and the mean first encounter time on a two-dimensional lattice with overlapping spatial domains.

The rest of the paper is organized as follows. The SIS model with a multi-layer network is introduced in section two. In section three, simulation results are presented and discussed in detail. The epidemic dynamics is analyzed in section four and conclusions are drawn in section five.

\section{The model}
\label{sec:model}
In the present model, each individual has two possible activities: acting as a random walker whose activity domain is restricted to a given local area and acting as a businessman whose activity is confined to interacting with his familiars. In order to model the evolutionary dynamics in local movement and global interaction (activating the prearranged linkage) respectively, we incorporate a multi-layer network with two subnetworks, which are called residential area and relationship network respectively throughout the paper, into the SIS model. The restricted residential area is common in territorial animals, where each individual is confined to local area and shares overlapping area with their neighbors\cite{giuggioli10,giuggioli11}. The relationship network is common in human society, which is a kind of relatively fixed person-to-person relation and has nothing to do with geographic distance\cite{newman}. Such relationship is usually modelled as a random network, a small world network or a scalefree network\cite{erdos10,watts10,cohen11,barabasi10}.

Similar to that in ref.\cite{giuggioli}. The residential area is a regular network with $L\times L$ sites, each of which has eight connected neighbor sites. The overall spatial domain is divided into $N_{local}=\frac{L}{d_x-\Delta d_x}\times \frac{L}{d_y-\Delta d_y}$ segregated spatial domains. The area of each domain is $s=d_x\times d_y$ and the overlapping area of two nearby domains is $\Delta s=d_y\times \Delta d_x$ or $d_x\times \Delta d_y$. There are total N ($\le$ $L\times L$) individuals staying on the sites of the residential area, each of whom can move along the link between the sites but is not allowed to walk out of the segregated spatial domain which is randomly allocated to each individual at the initial time.

The relationship network is a random network with N sites and $(N-1)\phi$ links on average for each site. Different sites denote different individuals and the links between sites denote social relations between the individuals. The prearranged interaction only takes place between the individuals with immediate connections. As the value of $\phi$ increases from 0 to unity, the relationship network changes from a disconnected network to a fully-connected network. Different from the original networked epidemic model, where the existence of the links between different individuals means the existence of the interactions between them. In the present model, the existence of these links only means the possibility of the interactions. Only when the links are activated, the interactions between the individuals really take place.

The reaction-diffusion process in the multi-layer network is as follows. There exists a timescale $\tau$ between prearranged interaction and local movement. At each time step, an individual i firstly makes his decision whether to interact with his familiar or go for a random walk. He makes prearranged interaction with probability $\tau$ and makes local movement with probability $1-\tau$. In the process of local movement, individual i firstly makes his decision whether to move or not with probability $v$. If he chooses "Yes", he randomly chooses a neighbor site from the total nine connected sites to stay on. If the chosen site is outside the local area, he does not move. If he chooses "No", he does not make a movement. After the movement decision has been made, for an infected individual, he becomes susceptible with probability $p_s$. For a susceptible individual, he becomes infected with probability $\frac{n_I}{n_s+n_I}p_I$, in which $n_s$ and $n_I$ are the number of susceptible individuals and infected individuals respectively and $n_s+n_I$ is the total number of individuals on the same site. In the process of prearranged interaction, individual i firstly chooses an individual j from his connected neighbors. Then, if individual i is an infected individual, he becomes susceptible with probability $p_s$. If individual i is a susceptible individual and individual j is an infected individual, individual i becomes infected with probability $p_I$. The updating is made asynchronously.

Therefore, as the value of timescale $\tau$ ranges continuously from 0 to unity, the present model changes from the territorial epidemic model to the random network epidemic model. The residential area and the relationship network are coupled by the individuals. The local epidemic dynamics and the global epidemic dynamics are coupled by the time scale $\tau$. For $0.5<\tau\le 1$, the global interaction process is faster than the local interaction process. For $\tau=0.5$, the global interaction process is the same as the local interaction process. For $0\le \tau<0.5$, the global interaction process is slower than the local interaction process.

\section{Simulation results and discussions}
\label{sec:results}
In the present model, we especially care about the coupled effects of local movement and prearranged interaction. In this section, the following parameters are given. The overlapping size is $\Delta d_x=\Delta d_y=1$. The overall spatial size is $L\times L=100\times 100$. The initial ratio of infected individuals is $\rho_I=0.05$. To reduce the influence of boundary effect, the segregated spatial domain is given as a square domain, that is, $d_x=d_y$.

\begin{figure}
\includegraphics[width=8cm]{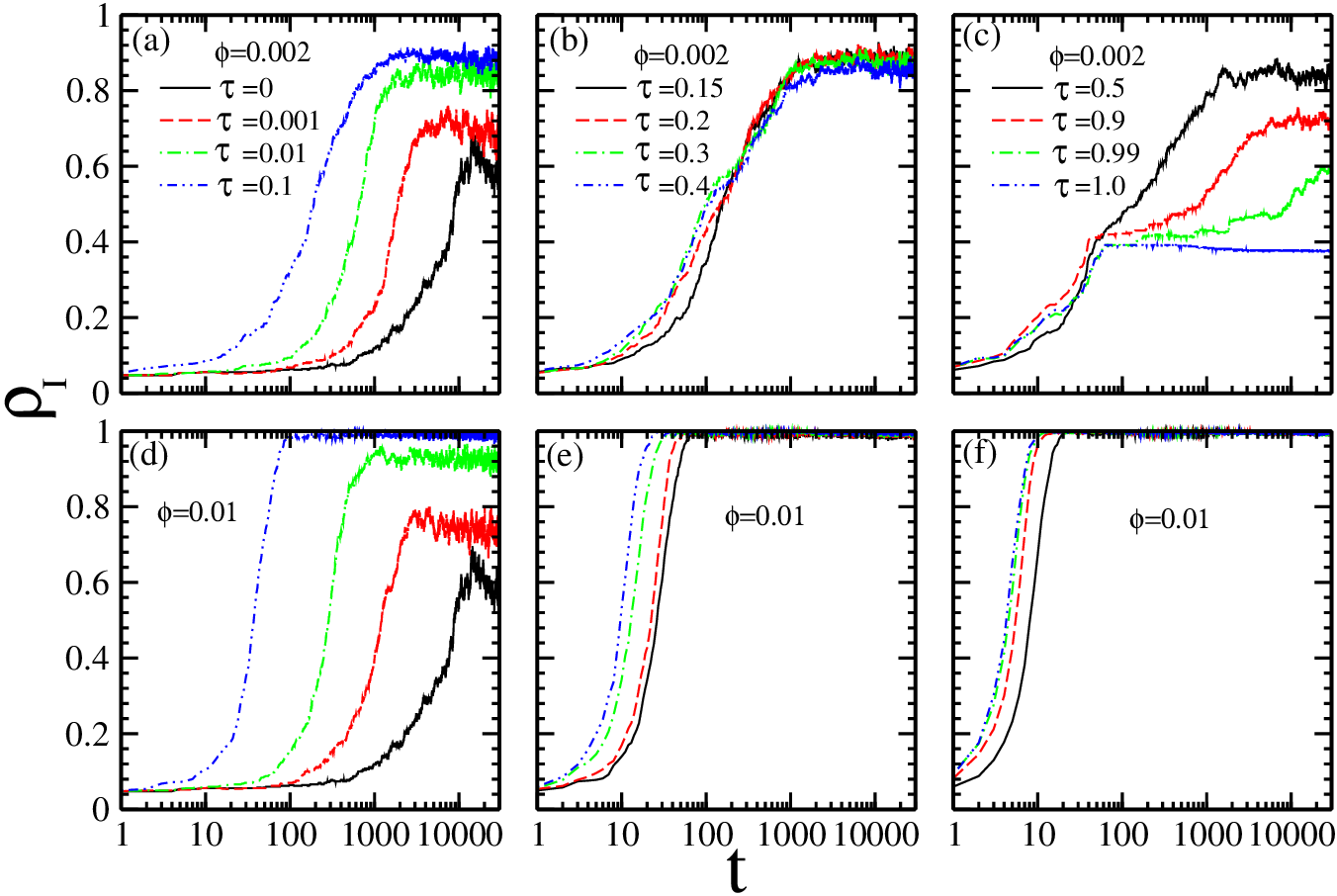}
\caption{\label{fig:epsart}The time-dependent ratio of infected individuals for a system with $N=500$, $p_I=1$, $p_s=0.001$, $d_x=d_y=11$, $v=0.05$, (a) $\phi=0.002$, $\tau$=0 (line), 0.001 (slash), 0.01 (slash dotted), 0.1 (slash dotted dotted; (b) $\phi=0.002$, $\tau$=0.15 (line), 0.2 (slash), 0.3 (slash dotted), 0.4 (slash dotted dotted; (c) $\phi=0.002$, $\tau$=0.5 (line), 0.9 (slash), 0.99 (slash dotted), 1 (slash dotted dotted); (d) $\phi=0.01$, $\tau$=0 (line), 0.001 (slash), 0.01 (slash dotted), 0.1 (slash dotted dotted; (e) $\phi=0.01$, $\tau$=0.15 (line), 0.2 (slash), 0.3 (slash dotted), 0.4 (slash dotted dotted; (f) $\phi=0.01$, $\tau$=0.5 (line), 0.9 (slash), 0.99 (slash dotted), 1 (slash dotted dotted). }
\end{figure}

In figure 1 we give the time-dependent ratio of infected individuals $\rho_I$ for different time scale $\tau$. In figure 1 (a) - (c), the ratio of prearranged linkage is $\phi=0.002$. In figure 1 (d) - (f), the ratio of prearranged linkage is $\phi=0.01$. Figure 1 (a) shows that, for $\tau$=0, which corresponds to the situation where the contact between different individuals only results from random walk, $\rho_I$ increases slowly with time and finally reaches an intermediate level of $\rho_I\sim0.6$. As $\tau$ increases from $\tau=0.001$ to $\tau=0.1$, the relaxation time decreases while the ratio of infected individuals in the final steady state increases with the rise of $\tau$. Similar results are also found in figure 1 (d). Therefore, within the range of $0<\tau<0.1$, the existence of shortcuts between segregated spatial domains has a great impact on the outbreak of the epidemic. The ratio of infected individuals in the final steady state is sensitive to the degree of activation of the prearranged linkage.

Different from the results in figure 1 (a), figure 1 (b) shows that,  within the range of $0.1<\tau<0.5$, the ratio of infected individuals in the final steady state is insensitive to the degree of activation of the prearranged linkage. As $\tau$ increases from $\tau=0.15$ to $\tau=0.45$, the ratio of infected individuals in the final steady state keeps the highest value of $\rho_I\sim 0.9$ and changes little with the rise of $\tau$. Similar results are also found in figure 1 (e).

Figure 1 (c) shows that, as $\tau$ increases from $\tau=0.5$ to $\tau=0.99$, the relaxation time increases while the ratio of infected individuals in the final steady state decreases with the rise of $\tau$. For $\tau$=1, which corresponds to the situation where the contact between different individuals only results from the prearranged linkage, the ratio of infected individuals in the final steady state is a low level of $\rho_I\sim0.37$. Different from the results in figure 1 (c), figure 1 (f) shows that, within the range of $0.5<\tau<1$, the ratio of infected individuals in the final steady state changes little with the rise of $\tau$. Such results indicate that, for a small value of $\phi$, which corresponds to the situation where the relationship network is disconnected, the existence of local movement can prompt the outbreak of the epidemic globally. For an intermediate or a large value of $\phi$, which corresponds to the situation where there is no disconnected individuals or disconnected clusters in the relationship network, the outbreak of the epidemic is determined by the structure of the relationship network.

As we keep an eye on the role of the prearranged linkage between different spatial domains in the outbreak of the epidemic, the role of activating the prearranged linkage in the present model is similar to the role of the existence of shortcuts in the two-dimensional small world epidemic model in ref.\cite{newman}. Only activating a small number of the prearranged linkage can prompt the outbreak of the epidemic globally.

\begin{figure}
\includegraphics[width=10cm]{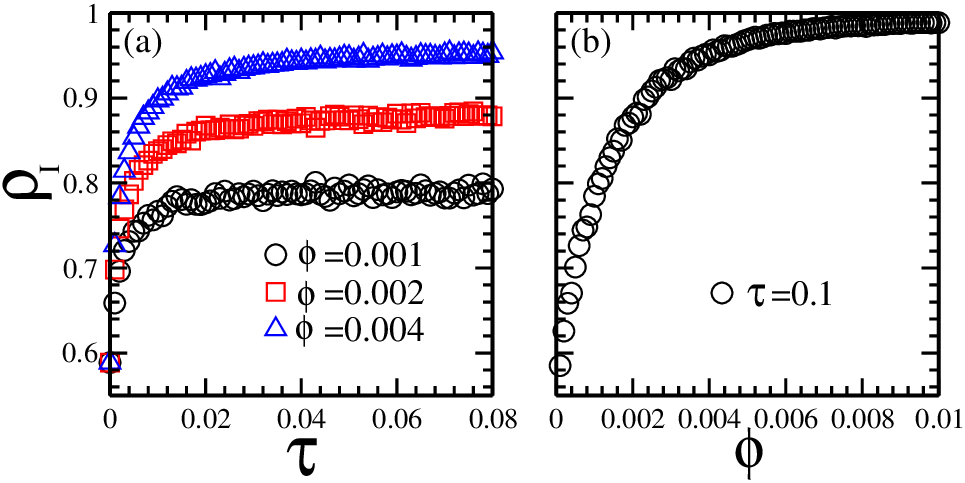}
\caption{\label{fig:epsart}The averaged ratio of infected individuals (a) as a function of timescale $\tau$ for the ratio of random connection $\phi$=0.001 (circles), 0.002 (squares), 0.004 (triangles) and (b) as a function of the ratio of random connection $\phi$ for $\tau=0.1$. The final data are obtained by averaging over 10 runs. In each run the relaxation time is $10^5$ time steps and the data are averaged over $10^4$ time steps. Other parameters are: $N=500$, $p_I=1$, $p_s=0.001$, $d_x=d_y=11$, $v=0.05$.}
\end{figure}

In order to find the quantitative relation between $\rho_I$ and $\tau$ for a given $\phi$ and the quantitative relation between $\rho_I$ and $\phi$ for a given $\tau$, in figure 2 (a), we give $\rho_I$ as a function of $\tau$ for $\phi=0.001, 0.002, 0.004$. In figure 2 (b) we give $\rho_I$ as a function of $\phi$ for $\tau=0.1$. Figure 2 (a) shows that, for a small $\phi=0.001$, within the range of $0<\tau<0.02$, $\rho_I$ increases quickly from $\rho_I\sim0.58$ to $\rho_I\sim0.78$. Within the range of $\phi>0.02$, $\rho_I$ changes little with the rise of $\tau$. For $\tau> 0$, an increase in $\phi$ leads to an overall rise of $\rho_I$. Figure 2 (b) tells us that, for a given $\tau=0.1$, only within the range of $0<\phi<0.008$, $\rho_I$ is quite sensitive to the change in $\phi$. Within the range of $\phi>0.008$, $\rho_I$ reaches its maximum value and changes little with the rise of $\phi$. Such results indicate that, in the scenario where the individuals are confined to segregated spatial domain, the existence of a small number of shortcuts can effectively lead to a rise of the ratio of infected individuals, which is similar to the result found in the small world network\cite{newman}.

\begin{figure}
\includegraphics[width=10cm]{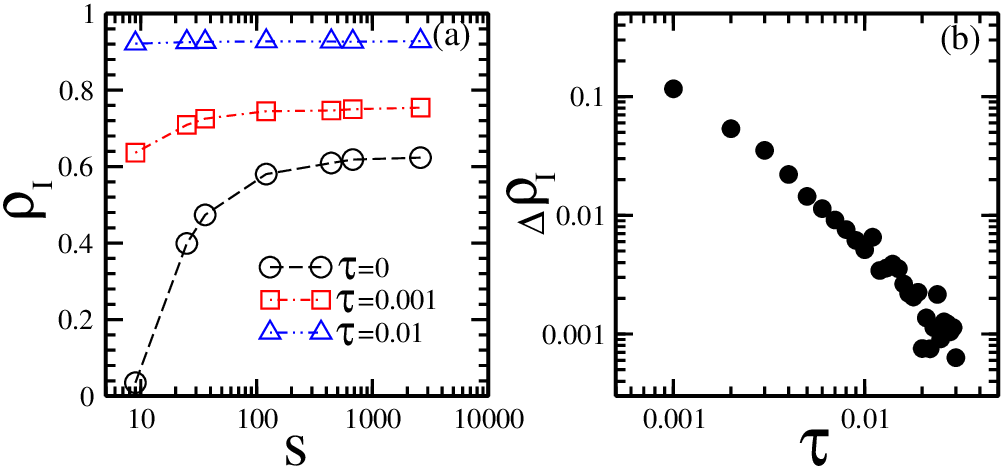}
\caption{\label{fig:epsart} (a) The averaged ratio of infected individuals as a function of the area s of segregated spatial domain for $\phi$=0.01, $\tau$=0 (circles), 0.001 (squares), 0.01 (triangles); (b) The change in the averaged ratio of infected individuals, $\Delta \rho_I=\rho_I^{s=2601}-\rho_I^{s=9}$, as a function of $\tau$. The final data are obtained by averaging over 10 runs. In each run the relaxation time is $10^5$ time steps and the data are averaged over $10^4$ time steps. Other parameters are: $N=500$, $p_I=1$, $p_s=0.001$, $v=0.05$.}
\end{figure}

The above results indicate that the evolutionary dynamics in the present model is determined by the mobility pattern and the connection pattern. To find out the relationship between local behavior and global behavior, in the following, the effects of three parameters on epidemic control are investigated respectively. The three parameters are: the area of segregated spatial domain s, the frequency of the interactions resulting from prearranged linkage $\tau$, and the ratio of segregated spatial domains with activity restriction $\zeta$ (called slow domain in the following).

Figure 3 (a) displays the averaged ratio of infected individuals $\rho_I$  as a function of $s$ for $\phi=0.01$ and $\tau=0, 0.001, 0.01$ respectively. For $\tau$=0, as s increases from s=9 to s=121, $\rho_I$ increases quickly from $\rho_I\sim 0.04$ to $\rho_I\sim 0.58$. As s increases from s=121 to s=2601, $\rho_I$ increases slowly and finally reaches its maximum value $\rho_I\sim 0.62$. Increasing $\tau$ leads to an overall rise of $\rho_I$ within the whole range of $9\le s\le 2601$. The smaller the value of s, the more sensitive of $\rho_I$ to the change in $\tau$.

The above results indicate that, by narrowing the segregated spatial domain (called activity domain restriction in the following), the epidemic spread may be inhibited, the effect of which depends on whether there exists global interaction or not. A higher frequency of global interaction will make activity domain restriction become less effective. Figure 3 (b) shows the relationship between the change of $\rho_I$, $\Delta \rho_I=\rho_I^{s=2601}-\rho_I^{s=9}$, and the time scale $\tau$. A log-log relation between $\Delta \rho_I$ and $\tau$, $\Delta \rho_I\sim a\tau^{-b}$, in which $a\sim 4.2\times 10^{-6}$ and $b\sim 1.53$, is found.

\begin{figure}
\includegraphics[width=10cm]{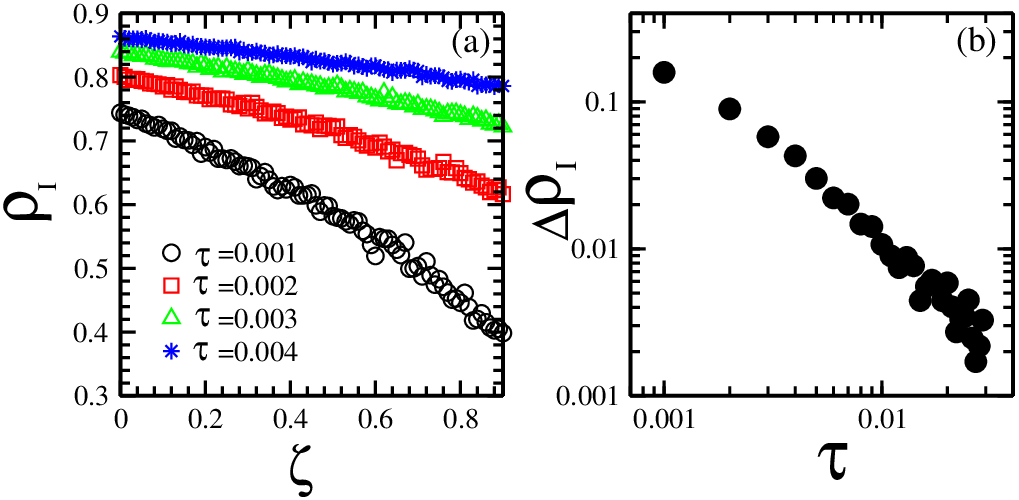}
\caption{\label{fig:epsart} (a) The averaged ratio of infected individuals as a function of the ratio of slow domains $\zeta$ for $\phi=0.01$ and $\tau$=0.001 (circles), 0.002 (squares), 0.003 (triangles), 0.004 (stars); (b) The change in the averaged ratio of infected individuals $\Delta \rho_I=\rho_I^{\zeta=0}-\rho_I^{\zeta=0.5}$ as a function of $\tau$ for $\phi$=0.01. The final data are obtained by averaging over 10 runs. In each run the relaxation time is $10^5$ time steps and the data are averaged over $10^4$ time steps. Other parameters are: $N=500$, $p_I=1$, $p_s=0.001$, $d_x=d_y=11$, $v=0.05$, $v_{slow}=0.1v$.}
\end{figure}

To find out whether local mobility restriction can effectively inhibit the global spread of epidemic, in figure 4 (a) we plot the averaged ratio of infected individuals as a function of the ratio of slow domain $\zeta$. In the slow domain, the moving probability is given as $v_{slow}=0.1v$. For $\tau=0.001$, as $\zeta$ increases from $\zeta\sim0$ to $\zeta\sim 0.9$, $\rho_I$ decreases continuously from $\rho_I\sim 0.74$ to $\rho_I\sim 0.4$. For a larger $\tau$, $\rho_I$ also decreases continuously with the rise of $\zeta$ but the slope of line becomes gentle.

The above results indicate that, local mobility restriction may inhibit the global spread of epidemic, the effect of which depends on whether there exists global interaction or not. A higher frequency of global interaction will make local mobility restriction become less effective. In figure 4 (b) we plot the relationship between the change of $\rho_I$, $\Delta\rho_I=\rho_I^{\zeta=0}-\rho_I^{\zeta=0.5}$, and the timescale $\tau$. A log-log relation between $\Delta \rho_I$ and $\tau$, $\Delta \rho_I\sim a'\tau^{-b'}$, in which $a'\sim 1.19\times 10^{-5}$ and $b'\sim 1.46$, is found. Such results indicate that the ratio of infected individuals is related to the average moving probability. An increase in the ratio of slow domain leads to a decrease in the average moving probability and then a decrease in the ratio of infected individuals.

The results in figure 3 and figure 4 indicate that, to make activity domain restriction and local mobility restriction become more effective on epidemic control, we should firstly slow down the frequency of global interaction. Or else, both the epidemic control methods can't play their proper role.

\section{Theoretical analysis}
\label{sec:analysis}
In order to get a clear picture of the microscopic characteristics of the propagation of the epidemic in the present model, in the following, we calculate the percolation threshold and the mean first encounter time on two-dimensional lattice with overlapping domains. Scaling laws for the epidemic threshold and the mean first encounter time are found.

\subsection{\label{subsec:levelA}relationship between epidemic threshold and activation of individual-based linkage}
The epidemic threshold is the critical value of infection probability, below which the ratio of infected individuals is small, while above which nearly all the individuals become infected. In a two-dimensional small world network, the epidemic threshold is found to be related to the connection range, the dimension of the lattice and the ratio of shortcuts\cite{newman,newman10}.

In the present model, without the individual-based linkage, which means $\phi=0$, the epidemic dynamics is similar to that in a regular network. Each individual can only interact with his nearest neighbors, those who are in the same segregated spatial domain or in the neighboring domain. In order to find the effect of the activation of the individual-based linkage on the evolutionary dynamics, in the following, the individuals are evenly distributed in the segregated spatial domains. Both the relationship between the epidemic threshold and the average number of individuals in each segregated domain and the relationship between the epidemic threshold and the moving probability are checked.

\begin{figure}
\includegraphics[width=10cm]{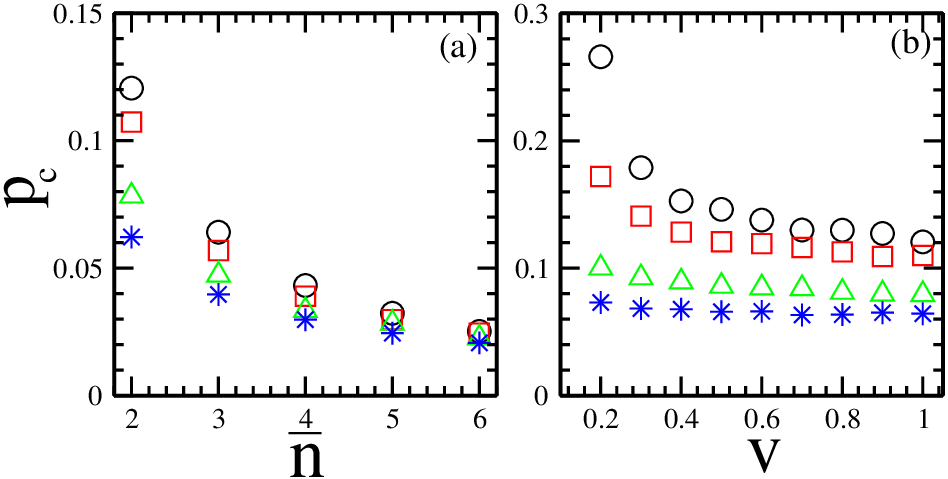}
\caption{\label{fig:epsart}The epidemic threshold (a)as a function of the number of individuals $\bar{n}$ in each segregated spatial domain for $\tau$=0(circles), 0.001(squares), 0.005(triangles), 0.01(stars);(b)as a function of the average moving probability $v$ for $\bar n$=2 and $\tau$=0(circles), 0.001(squares), 0.005(triangles), 0.01(stars); The final data are obtained by averaging over 10 runs. In each run the relaxation time is $5\times10^4$ time steps and the data are averaged over $10^3$ time steps. Other parameters are: $p_s=0.001$, $d_x=d_y=11$, $\phi=0.01$.}
\end{figure}

In figure 5 (a) we give the epidemic threshold as a function of the number of individuals $\bar{n}$ in each segregated spatial domain for the ratio of the individual-based linkage $\phi=0.01$ and different timescale $\tau$. For $\tau=0$, which corresponds to the situation where all the individuals are confined to local domains, as $\bar{n}$ increases from $\bar{n}=2$ to $\bar{n}=6$, the epidemic threshold $p_c$ ranges from $p_c\sim 0.121$ to $p_c\sim 0.025$. An increase in $\tau$ leads to an overall decrease in $p_c$. A scaling law for $p_c\sim\frac{1}{\bar n^{\alpha}}$ is found. As $\tau$ increases from $\tau=0$ to $\tau=0.01$, the value of $\alpha$ decreases from $\alpha\sim 1.418$ to $\alpha\sim 0.998$.

In figure 5 (b) we give the epidemic threshold as a function of the average moving probability $v$ for $\bar n=2$, $\phi=0.01$ and different $\tau$. For $\tau=0$, as the value of $v$ ranges from $v=0.2$ to $v=1$, $p_c$ decreases from $p_c\sim 0.266$ to $p_c\sim 0.121$. An increase in $\tau$ leads to an overall decrease in $p_c$. A scaling law for $p_c\sim\frac{1}{v^{\beta}}$ is found. As $\tau$ increases from $\tau=0$ to $\tau=0.01$, the value of $\beta$ ranges from $\beta\sim 0.434$ to $\beta\sim 0.077$.

Comparing the results in figure 1 (a) with the results in figure 5 (a) and (b), we find that the effect of activating the prearranged linkage on the change of the epidemic threshold is similar to the effect of activating the prearranged linkage on the change of the ratio of the infected individuals in the final steady state. The smaller the value of the epidemic threshold, the higher the value of the ratio of the infected individuals in the final steady state. For a small value of timescale $\tau$, both the epidemic threshold and the ratio of the infected individuals in the final steady state are sensitive to the change of $\tau$. Such a result is similar to what has been found in the small world epidemic model\cite{newman,newman10}.

The above results can be understood as follows. For $\tau$=0, as each individual is confined to his local domain, a longer separation leads to a higher epidemic threshold and a smaller ratio of the infected individuals in the final steady state. For $\tau\ne 0$, the activation of a small number of shortcuts produces a shorter separation, which leads to a lower epidemic threshold and a larger ratio of the infected individuals in the final steady state.

\subsection{\label{subsec:levelB}Relationship between the mean first encounter time and the size of segregated spatial domain on two dimensional lattice}

The propagation speed of the epidemic is related to the pairwise interaction between an infected individual and a susceptible individual. If the mean first encounter time between different individuals is small, it is much easier for the epidemic to break out. On one-dimensional lattice with segregated spatial domains, it has been found that the mean first encounter time is related to the size of the segregated domain $d_x$  and the size of the overlapping area $\Delta d_x$\cite{giuggioli}. In the present model, the individuals are confined to two-dimensional lattice with segregated spatial domains. By calculating the mean first encounter time, we can find out the relationship between the encounter probability and the variables of the size of the segregated domain $s$ and the average number of individuals in each segregated domain $\bar{n}$. The microscopic characteristics of the epidemic on a two-dimensional lattice with segregated spatial domains can be understood.

\begin{figure}
\includegraphics[width=10cm]{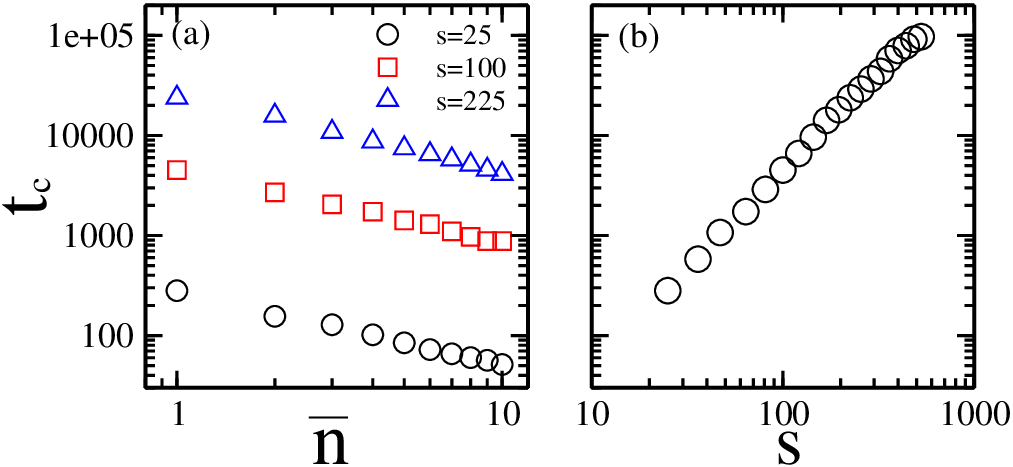}
\caption{\label{fig:epsart}The mean first encounter time (a) as a function of the average number of individuals in each segregated spatial domain for $s$=25(circles), 100(squares), 225(triangles) and (b) as a function of the area of segregated spatial domain for $\bar n=1$. Other parameters are $\Delta d_x=\Delta d_y=1$, $v=1$, $\phi$=0, $\tau$=0. Final results are averaged over 50 runs.}
\end{figure}

In figure 6 (a) we plot the mean first encounter time as a function of the average number $\bar{n}$ of individuals in each segregated spatial domain for $\phi=0$. For simplicity and refraining from the border effect, here we only consider the case where the domain is satisfied with the condition $d_x=d_y$. It is observed that, for $d_x=d_y=5$, as $\bar{n}$ increases from $\bar{n}=1$ to $\bar{n}=10$, $t_c$ decreases from $t_c\sim 280$ to $t_c\sim 50$. Increasing $d_x$ ($d_y$) leads to an overall rise of the value of $t_c$. A scaling law for $t_c\sim\frac{1}{\bar{n}^\gamma}$, in which $\gamma\sim0.73$, is found. As the value of $d_x$ changes from $d_x=5$ to $d_x=15$, the value of $\gamma$ changes little. In figure 6 (b) we plot the mean first encounter time as a function of the size of the segregated spatial domain for $\bar n$=1. A scaling law for $t_c\sim\frac{1}{s^k}$, in which $k\sim1.96$, is found.

Comparing the results in figure 3 (a) with the results in figure 6 (a) and (b), we find that a decrease in $\rho_I$ by narrowing the segregated spatial domain should result from the competition of the following two processes: a decrease in the area of the segregated spatial domain and a decrease in the average number of individuals in each domain. In the present model, the population size and the spatial size are fixed. A decrease in the size of the segregated spatial domain leads to an increase in the total number of the segregated spatial domains and a decrease in the average number of individuals in each domain. From the simulation results in figure 6 (a) and (b) we find that a decrease in the size of the segregated spatial domain leads to a decrease in the mean first encounter time while a decrease in the average number of individuals in each segregated spatial domain leads to an increase in the mean first encounter time. In the present model, the decrease of the mean first encounter time resulting from the decrease of the size of the segregated spatial domain should be slower than the increase of the mean first encounter time resulting from the decrease of the number of individuals in each segregated spatial domain. Therefore, the coupling of the above two contradictory processes leads to the occurrence of the macroscopic characteristics of the outbreak of the epidemic in figure 3 (a), that is, the smaller the size of the segregated spatial domain, the lower the ratio of the infected individuals in the final steady state.

\section{Summary}
\label{sec:summary}
By incorporating segregated spatial domains and individual-based linkage into the SIS model, we investigate the coupled effects of local movement and global interaction on contagion. The role of activating the prearranged linkage between different spatial domains in the outbreak of the epidemic is especially concerned. Compared with the scenario where only random walk exists, the existence of shortcuts between different spatial domains leads to a wider spread of the epidemic. The effects of home range restriction and mobility restriction on epidemic control are extensively investigated. When the global interaction is restricted, both narrowing the segregated spatial domain and reducing local mobility can effectively curb the spread of the epidemic. Activating the prearranged linkage makes the epidemic control methods become less effective.

A heuristic analysis indicates that the macroscopic characteristics of the propagation of the epidemic in the present model is closely related to the microscopic characteristics of the individuals moving on two-dimensional lattice. Activating the shortcuts between different spatial domains lowers the epidemic threshold effectively and therefore prompt the outbreak of the epidemic. Narrowing the segregated spatial domain leads to a decrease in the area of random walk and a decrease in the average number of individuals in each area simultaneously. The coupling of the two processes results in an increase in the mean first encounter time and therefore the effectiveness of activity restriction on epidemic control.

In the future, the coupled effects of mobility patterns and interaction patterns on the evolutionary dynamics will be extensively studied in reaction-diffusion systems and game models. The evolutionary dynamics of one to many interactions in similar coupled systems will be investigated in the near future.

\section*{Acknowledgments}
This work is the research fruits of National Natural Science Foundation of China (Grant Nos. 71371165, 11175079, 71471161, 71171176, 71471031, 71171036, 71072140), Research Project of Zhejiang Social Sciences Association (Grant No. 2014N079), Collegial Laboratory Project of Zhejiang Province (Grant No. Z201307), Visiting Scholar Project of Teacher Professional Development of Zhejiang Province(Grant No. FX2014057), Jiangxi Provincial Young Scientist Training Project (Grant No. 20133BCB23017). The major project of the National Social Science Foundation of China (Grant No.12 ZD067, 14AZD089), the Program of Distinguished Professor in Liaoning Province (Grant No.[2013]204), the Project of Key Research Institute of Humanities and Social Sciences by Department of Education of Liaoning Province (Grant No. ZJ2013037), the Program of Innovative Research Team and Discipline Support Plan in Dongbei University of Finance and Economics(Grant No. DUFE2014T01, XKK-201401).





\bibliographystyle{model1-num-names}



\end{document}